\documentclass[iop,apj,preprint]{emulateapj}

\def\subtitle{
  \vspace*{-12mm}
  \noindent
  {\scriptsize {\sc \@journalinfo} \\
        Preprint typeset using \LaTeX\ style emulateapj v.\ \emulateapj@ver}
}
\submitted{Accepted version June 2, 2021}
\journalinfo{\@ApJ}

\usepackage{lineno}
\usepackage{amsmath,amssymb,amstext}

\usepackage[breaklinks,colorlinks,citecolor=blue,linkcolor=magenta]{hyperref}

\usepackage{xcolor}

\usepackage[all]{hypcap} 

\shorttitle{Explosive ejections generated by gravitational interactions}
\shortauthors{Rivera-Ortiz, Rodr\'iguez-Gonz\'alez, Cant\'o \& Zapata}

\begin{document}
\title{Explosive ejections generated by gravitational interactions}
\author{P. R. Rivera-Ortiz\altaffilmark{1}, A. Rodr\'iguez-Gonz\'alez\altaffilmark{2}, J. Cant\'o\altaffilmark{3}, and Luis A. Zapata\altaffilmark{4}}
\affiliation{$^1$ Univ. Grenoble Alpes, CNRS,
Institut de Planétologie et d’Astrophysique de Grenoble (IPAG), 38000 Grenoble, France}
\affiliation{$^2$ Instituto de Ciencias Nucleares, 
Universidad Nacional Aut\'onoma de M\'exico, 
Ap. 70-543, 04510 D.F., M\'exico}
\affiliation{$^3$Instituto de Astronom\'{\i}a, Universidad
Nacional Aut\'onoma de M\'exico, Ap. 70-264, 04510 D.F., M\'exico
\affiliation{$^4$Instituto de Radioastronom\'ia y Astronom\'{\i}a, Universidad
Nacional Aut\'onoma de M\'exico, Apdo. Postal 3-72 (Xangari), 58089 Morelia, Michoac\'an, M\'exico}}

\begin{abstract}
During the fragmentation and collapse of a molecular cloud, it is expected to have close encounters between (proto)stellar objects that can lead to the ejection of a fraction of them as runaway objects. However, the duration and the consequences of such encounters perhaps are small {\bf such} that there is no direct evidence of their occurrence. As a first approximation, in this work, we analytically analyze the interaction of a massive object that moves at high velocity into a cluster of negligible mass particles with an initial number density distribution  $\propto R^{-\alpha}$. We have found that the runaway conditions of the distribution after the encounter are related to  the mass and the velocity of the star and the impact parameter of each particle to the stellar object. Then, the cluster particles are gravitationally accelerated by the external  approaching star, destroying the cluster and the dispersion and velocities of the particles have explosive characteristics. We compare this analytical model with several numerical simulations and finally, we applied our results to the Orion Fingers in the Orion BN/KL region, which show an explosive outflow that could be triggered by the gravitational interaction of several (proto)stellar objects.

\end{abstract}

%
%
%


\section{Introduction}
\label{sec:intro}

During the last decades the star formation processes had been extensively studied, which has led to a general comprehension of the stages that drive a molecular cloud to collapse, to fragment, and to form new stars  \citep{MO07, K14}arriving to form stellar clusters in a star forming region and the star forming efficiency and the conditions to create and preserve a gravitational bound have to be considered  \citep{KETAL14}. This problem has been analyzed by \cite{F10} and \cite{kr12}. In any case, and even when a forming cluster is bounded, the local conditions could eject some stars as a result of close gravitational interactions from a multiple star system \citep{IETAL18}. These encounters are expected, at least in the densely populated regions, but there is not any observational direct evidence  of them. Then, the duration, characteristics and their influence in the interstellar medium of such encounters are still unknown.

It has been suggested recently that the explosive outflows reviewed by
\cite{BETAL17} may be produced as a consequence of those kind of close encounters. The closest of these explosive outflows is Orion BN/KL, which may have been produced by a close gravitational interaction of several protostellar objects (\citealt{BN67}, \citealt{kl67}). This outflow shows a characteristic filamentary structure emitting in H$_{2}$ known as {\it Orion fingers} with a kinetic energy of around $10
^{47}$-$10^{48}$erg distributed almost isotropically around a common center, that is the same common origin for the runaway stellar objects BN, x and the binary I, with masses around 10M$_\odot$ (\citealt{BETAL2020}). However, motivated by the Orion BN/KL morphology, \citet{AB93} proposed that this outflow could have an explosive origin were some dense clumps were ejected at very high velocities that interact with the surrounding environment and the resulting wakes are the actual fingers. At the tip of such fingers there is highly excitation emission, eg FeII, produced by the high velocities of these  bullets, that have proper motions as large as 300~km~s$^{-1}$ (\citealt{AB93}, \citealt{CPHD06}, \citealt{NETAL07}). \citet{ZETAL09} reported a set of streamers emitting in CO J=2$\to$1, that are related to the fingers, and that follow a {\it Hubble law} that is a signature of other explosive outflows like DR21 \citep{ZETAL13} and G5.89 (see \citep{ZETAL19} and  \citep{ZETAL20}).

 The problem of a star ejected from a stellar cluster or the evaporation of members from a stellar cluster has been previously investigated by several authors, including \citet{B61}, \citet{K66}, \citet{PS12} and \cite{WETAL19}. They analyzed the close encounters of cluster members, which lead to form close binaries and another star takes the excess of energy as kinetic energy. Nevertheless, this mechanism does not take into account the formation of explosive outflows (reviewed by 
\cite{BETAL17}). It has been suggested that a close encounter of an already runaway star disrupt into a stellar cluster. Particularly, \citet{BETAL15} proposed that an external runaway star disrupted a forming stellar cluster, and the debris of this disruption is the explosive outflow from Orion BN/KL.

In order to explain the explosive outflow of Orion BN/KL, \citet{BETAL15} proposed a qualitative model to describe the close interaction of a forming stellar cluster that is disturbed by  an external high velocity and massive object destroying the former cluster through unstable orbits, then producing smaller bounded systems and releasing the excess of energy in the environment gas through several close interactions of the involved objects expelling gas clumps that could form the filamentary structure  of the explosive outflow. In the literature exists evidence of high velocity (up to several hundreds~km~s$^{-1}$)  stellar objects, such as runaway neutron stars \citep{IETAL18} and {\bf \citet{DJETAL20} } analyzed a sample of runaway stars with masses from 10 to 60 M$_\odot$ and velocities from 20 to 200 km s$^{-1}$ which makes plausible that a runaway star could impact and disrupt a cluster. Also, \citet{BETAL2020} performed and reviewed several hydrodynamical simulations that investigate the probability of a system of 4 stellar objects to form a system of runaway stars as the one observed in Orion BN/KL.  Nevertheless, the mechanism that triggered explosive outflows has not been deeply explored, since it cannot explain the acceleration and dispersion of the clumps that lead the fingers compared with the observed velocities. Also, the repeated encounters seems to be against the idea of a single explosive event.

 As \citet{BETAL15} proposed, the Orion BN/KL could have suffered an interaction between a forming stellar cluster and an external stellar object, disrupting the cluster members, but the origin of the high velocity features known as the H$_2$ Orion fingers and the CO streamers, that have velocities the order of several hundreds of km s$^{-1}$, is still unknown.   Then, explaining the ejection of the residual clumps from a disrupting cluster can lead to a better understanding of the explosive outflows.  The nature of this cluster is actually not known and its geometry, size or mass are a matter of debate. \citet{NETAL07} found an inferior limit for the leading clumps of the fingers of $10^{-5}M_\odot$ and 
 according to \citet{ROETAL19}, \citet{ROETAL19b} (hereafter RO19a and RO19b, respectively) and \citet{dzo20}, the initial mass of each clump is in the planetary mass scale, so it could be reasonable to think that they come from the young proto-planetary disk of one of the stellar objects. Also,  according to \citet{MoNe12} and \citet{BIETAL12} the number, mass and distribution of the planetary embryos in the proto-planetary disks are related to the time evolution of a star.
These clumps would suffer a similar interaction as the one described in \citet{BETAL15}, such as a Rutherford dispersion originated by a gravitational force, affecting them drastically and changing their dynamical properties via one gravitational impulse \citep{SETAL70} produced by the passage of a massive object. 
As a first approximation, we describe the interaction of a fast massive star with a cluster of negligible mass objects.

In this paper we analyze the gravitational effect of a massive star on a cloud of smaller objects 
 as the possible origin of the explosive outflows described before. In Section \ref{sec:model} we propose the analytical tools to describe the interaction, in Section \ref{sec:simulations} we describe the N-Body simulations used to calibrate the analytical model. We propose some interpretations applied to Orion BN/KL in Section \ref{sec:Orion} and finally, we included our conclusion in Section \ref{sec:conclusions}.

\section{Model}
\label{sec:model}

 In this paper we address the problem of the gravitational collision between a cluster of cloudlets of negligible mass and a massive object with mass $M$, moving at high velocity $v_0$. 

\subsection{Static massive particle}

 We first consider the motion of one of these cloudlets in the frame of reference of the colliding object (the system $xy$, see Fig \ref{fig:model}). In this frame of reference, the cloudlet approaches the massive object with a velocity $v_0$ (at infinity), along an initial straight trajectory with an impact parameter $\xi$.  Considering that initial gravitational interaction is small enough to be neglected, the mechanical energy of the incident particle is positive (equal to $v_0^2/2$ per unit mass) the trajectory is a hyperbole. 

We define the gravitational radius $\xi_0$,

\begin{equation}
    \xi_0=\frac{GM}{v_0^2},
\end{equation}
 and the characteristic time $t_0$,
 
\begin{equation}
    t_0=\frac{\xi_0}{v_0}.
\end{equation} 
 
Using these quantities, we adimensionalise distances with $\xi_0$, velocities with $v_0$ and times with $t_0$.

Then, the hyperbolic trajectory of a cloudlet, in polar coordinates, is given by (\citet{CA89}; see also \citet{CETAL11}),
 
\begin{equation}
r=\frac{\xi'^2}{1-\cos \theta  + \xi' \sin \theta},
\label{eq:rtheta}
\end{equation} 

where $r$ and $\xi'$ are the dimensionless distance to the particle and impact parameter, respectively. 

In Equation (\ref{eq:rtheta}), the angle $\theta$ is measured {\bf with} respect to $+\hat{x}$ direction (see Fig. (1)). From this equation, we can observe that for $r\rightarrow \infty$, the cloudlet moves along incoming and outcoming asymptotic straight lines, which are given by the angles $\theta$ equal to 0  and $\theta_{\infty}$. This last sentence can be written as:
\begin{equation}
    \cos\theta_{\infty}=\frac{1-\xi'^2}{1+\xi'^2}.
    \label{eq:thetainf}
\end{equation}
We set the scattering angle, i.e. the angle between the incoming and outcoming directions of the movement, as 
$\alpha_m=\theta_\infty-\pi$ (see Fig. (1)). From Equation (\ref{eq:thetainf}), we obtain:
\begin{equation}
    \cos\alpha_m=\frac{\xi'^2-1}{\xi'^2+1}.
    \label{eq:alpham}
\end{equation}

Our next step is to obtain the $x-$ and $y-$ components of the cloudlet's velocity as a function of $\theta$ and $\xi'$. The dimensionless velocity can be written in polar coordinates as:
\begin{equation}
    \mathbf{u}=u_r \hat{r}+u_\theta \hat{\theta}=\frac{dr}{d\tau} \hat{r}+ r \frac{d\theta}{d\tau} \hat{\theta}= \frac{d\theta}{d\tau}\bigg(\frac{dr}{d\theta}\hat r + r \hat \theta\bigg),
    \label{eq:upolar}
\end{equation}
where $\tau$ is the dimensionless time. In this problem, the angular momentum is conserved, which is given by:
\begin{equation}
    \xi'=r u_\theta = r^2 \frac{d\theta}{d\tau}
\label{eq:angmom}
\end{equation}
Then, we finally obtain $u_x$ and $u_y$, putting Eqs (\ref{eq:rtheta}) and (\ref{eq:angmom}) in Eq.(\ref{eq:upolar}), and writting the directions $\hat r$ and $\hat \theta$ as a functions of $\hat x$, $\hat y$, and $\theta$:
\begin{equation}
   \mathbf{u}=u_x \hat x + u_y \hat y=-\frac{1}{\xi'}(\xi'+\sin\theta) \hat{x} + \frac{1}{\xi'}(1-\cos\theta) \hat y
   \label{eq:vxvy}
\end{equation}

Equations~(\ref{eq:vxvy}) can be used to obtain the velocity module $u=\sqrt{u_x^2+u_y^2}$ and, after some algebra we obtain the mechanical energy conservation equation,

\begin{equation}
    \frac{1}{2}u^2-\frac{1}{r}=\frac{1}{2}
\end{equation}

Besides, we find a relationship between $d\tau$ and $d\theta$, combining Eqs.(\ref{eq:rtheta}) and (\ref{eq:angmom}), which is given by:
\begin{equation}
    d\tau = \frac{\xi'^3}{(1-\cos\theta+\xi'\sin\theta)^2} d\theta
\end{equation}

which can be integrated to give


\begin{equation}
\tau+c=\ln[1+\xi'\cot \frac{\theta}{2}] + \frac{1+\xi'^2}{2[1+\xi'\cot \frac{\theta}{2}]}-\frac{\xi'}{2}\cot \frac{\theta}{2},
\end{equation}

with $c$ as an integration constant. By choosing $\tau=0$ for an angle $\theta=\pi$ in the limit $\xi'=0$ we find that $c=\frac{1}{2}$. 

Defining $\mu=[1+\xi'\cot \frac{\theta}{2}]$ the relation between the position angle, e.g. the angle $\theta$ and the time $\tau$ is given by

\begin{equation}
    \tau=\ln[\mu]+\frac{1+\xi'^2}{2\mu}-\frac{\mu}{2}.
    \label{eq:time}
\end{equation}

For an angle $\theta=\pi$, according to Equations~(\ref{eq:rtheta}) and (\ref{eq:time}) the particle crosses the $x$-axis ($\theta=\pi$) at a time and position that depend on the impact parameter, such that,
\begin{eqnarray}
\tau_*&=&\frac{\xi'^2}{2} \label{eq:t-axis}\\
x_*&=&-\frac{\xi'^2}{2} \label{eq:r-axis} .
\end{eqnarray}

Note that, correctly, for $\xi=0$, $\tau_*=x_*=0$.

\subsection{Moving massive particle}

\begin{figure}
    \centering
    \includegraphics[width=0.9\columnwidth]{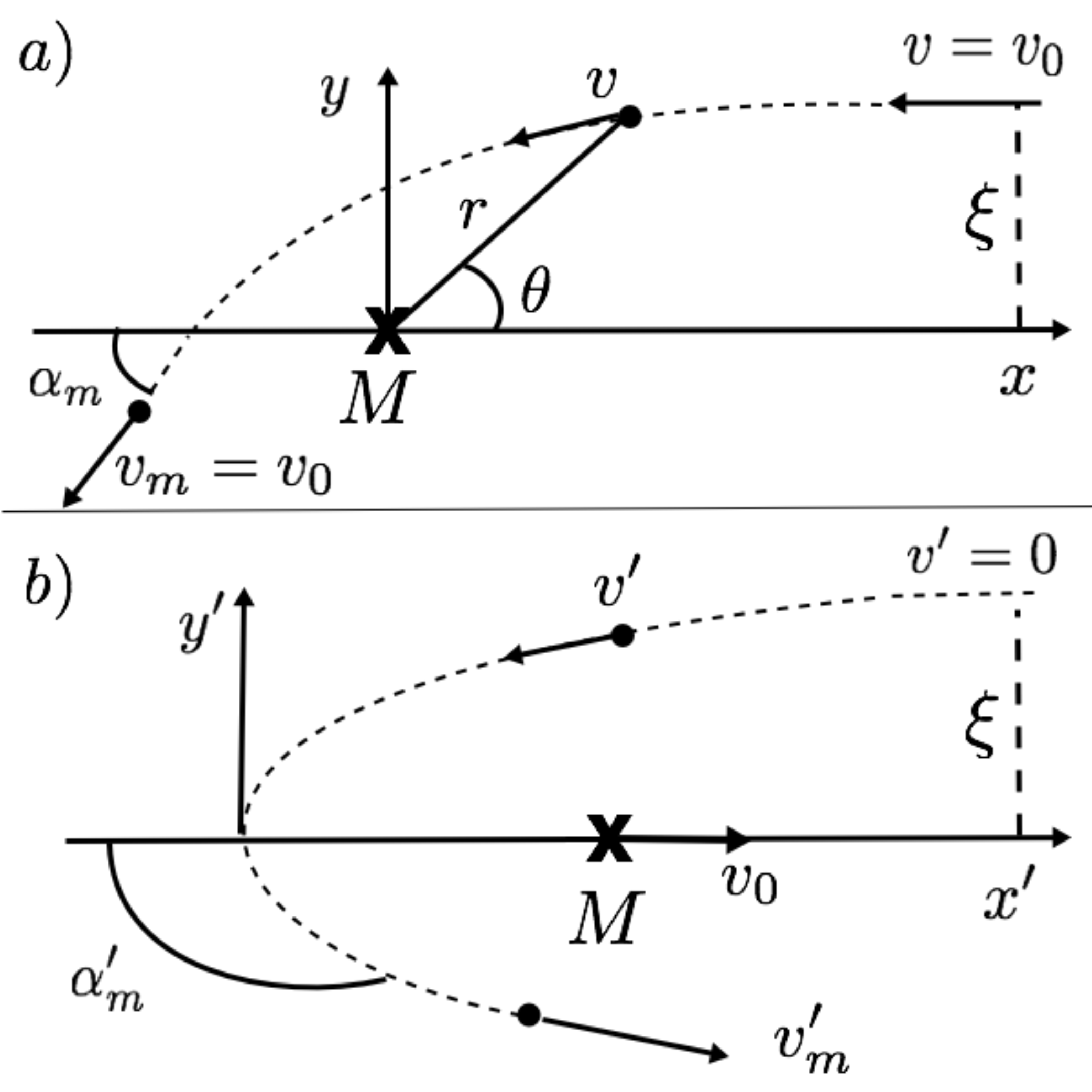}
    \caption{Gravitational interaction between a mass $M$ moving with velocity $v_0$ and a massless cloudlet at rest. a)In the frame of reference of the mass $M$, the cloudlet initially moves with velocity $v_0$ at infinity, along a straight trajectory with impact parameter $\xi$. After the collision, the cloudlet deviates from its initial trajectory by an angle $\alpha_m$ and regains a velocity $v_0$ equal to its initial velocity. b) In the frame of reference of the cluster, the cloudlet is initially at rest while the star is moving with velocity $v_0$. The cloudlet is accelerated and is thrown away at an angle $\alpha_m'$ with velocity $v_m'$ (see text) }
    \label{fig:model}
\end{figure}

In this subsection, we adopt the frame of reference where both the observer and the cloudlets are at rest with respect to each other, and the massive particle moves with velocity $v_0$ (see Fig. (\ref{fig:model})). In this frame of reference, the cloudlet has coordinates 

\begin{equation}
x'=x+\tau, \quad \; \; y'=y,
\end{equation}
 
 and velocities,
 
\begin{equation}
    u_x'=u_x+1, \quad \; \; u_y'=u_y.
\end{equation}

From this last Equation and considering Equation (\ref{eq:vxvy}) we find that the velocity is given by

\begin{equation}
\label{eq:vp2}
    u'=\frac{[2(1-\cos\theta)]^{1/2}}{\xi'}.
\end{equation}

Then, the direction in which the cloudlet moves after interacting with the massive particle and far away from it, i.e. for $\theta=\theta_\infty$ (see Equation (\ref{eq:thetainf}), forms an angle $\pi-\alpha'_m$ with respect to $\hat{x}$ (see Panel b of Fig. (\ref{fig:model})). This angle is obtained by doing

\begin{equation}
    \cos(\pi-\alpha'_m)=-\cos \alpha'_m=\frac{u'_x}{u'}=\frac{-\sin \theta_\infty}{\sqrt{2(1-\cos \theta_\infty)}}.
    \label{eq:deviation}
\end{equation}

Considering Equation (\ref{eq:thetainf}) into Equation (\ref{eq:deviation}), we get

\begin{equation}
\label{eq:thetam}
    \alpha'_m=\arccos\left(-\frac{1}{(1+\xi'^2)^{1/2}}\right).
\end{equation}

Finally, the position $x_*'$ when the particle crosses the $x$-axis is

\begin{equation}
    x_*'=x_*+\tau_*=0,
\end{equation}

which is independent of $\xi$ (see Equations (\ref{eq:r-axis}) and (\ref{eq:t-axis}))and therefore, every particle crosses the $x$-axis in the same point although at different time.

Combining Equation~(\ref{eq:vp2}) and (\ref{eq:thetam}) we obtain:

\begin{equation}
     u=\frac{2}{(1+\xi'^2)^{1/2}}
\end{equation}

Using the dimensional expressions we have:

\begin{equation}
    \label{eq:vm2}
     v=\frac{2 v_0}{(1+\xi'^2)^{1/2}}
\end{equation}

\subsection{The velocity distribution}

Consider a cluster of radius $\xi_c$ containing $N_T$ clumps. The density distribution is

\begin{equation}
    n(r)=Ar^\alpha.
\end{equation}

Then,

\begin{equation}
    N_T= \int_0^{\xi_c}4\pi r^2 n(r)dr =\frac{4\pi A}{3+\alpha}\xi_c^{3+\alpha}
\end{equation}

{\bf for $\alpha\neq -3$ } and thus,

\begin{equation}
    A=\frac{(3+\alpha)N_T}{4\pi \xi_c^{3+\alpha}}
\end{equation}

Let $dN(\xi)$ be the number of clumps with impact parameter between $\xi$ and $\xi + d\xi$. We define the distribution $g(\xi)$ of impact parameters such that

\begin{equation}
    dN(\xi)=N_Tg(\xi)\, d\xi.
    \label{eq:dN1}
\end{equation}

Clearly 
\begin{equation}
    \int_0^{\xi_c} g(\xi)d\xi=1
\end{equation}

where $\xi_c$ is the maximum impact parameter. 

\begin{figure}
    \centering
    \includegraphics[width=0.7\columnwidth]{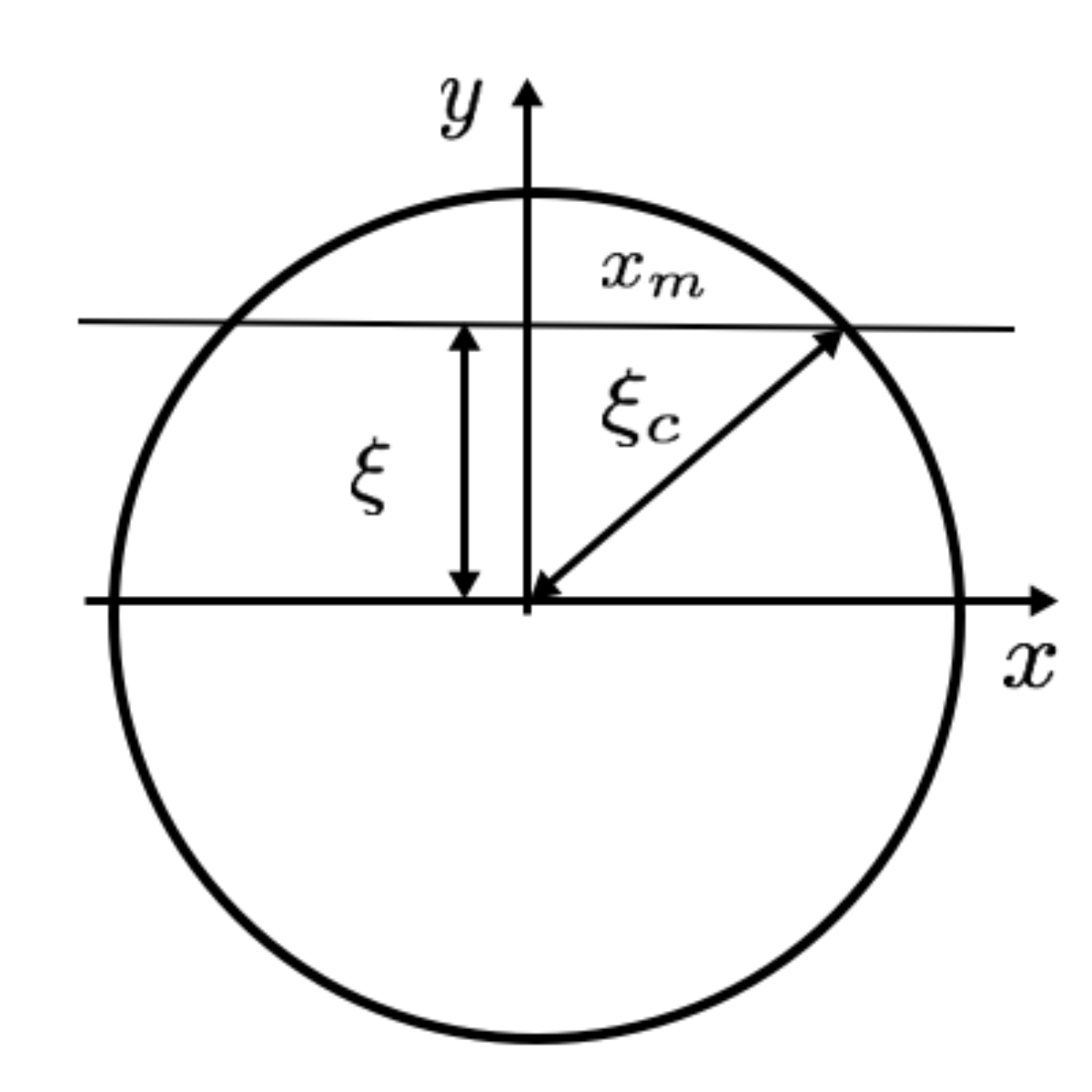}
    \caption{Diagram of the cloudets cluster with size $\xi_c$. The $x$-axis is parallel to the direction of motion of the massive particle at infinity and, for a given impact parameter $\xi$ there is a distribution function $g(\xi)$ that accounts for the fraction of cloudlets that are between $\xi$ and $\xi+d\xi$ in the limits $-x_m$ and $x_m$.}
    \label{fig:xm}
\end{figure}

From Fig. (\ref{fig:xm}) {\bf we find}

\begin{equation}
    dN(\xi)=2\pi \xi d\xi \left( 2\int_0^{x_m} n(x)\, dx\right)
    \label{eq:dN}
\end{equation}

where $x_m=(\xi_c^2-\xi^2)^{1/2}$ and $n(x)=Ar^\alpha=A(\xi^2+x^2)^{\alpha/2}$

Let 

\begin{equation}
    \int_0^{x_m} n \, dx =A \int_0^{x_m} (\xi^2+x^2) ^{\alpha/2} \, dx=AI_\alpha
\end{equation}

where
\begin{equation}
 I_\alpha=\int_0^{x_m} (\xi^2+x^2) ^{\alpha/2} \, dx.
\label{eq:Ialpha}
\end{equation}
Then, from Equations (\ref{eq:dN}) and (\ref{eq:Ialpha}) we can express $g(\xi)$ as

\begin{equation}
    g(\xi)=4\pi A \xi I_\alpha/N_T=\frac{(3+\alpha)\xi}{\xi_c^{3+\alpha}}I_\alpha,
\end{equation}

where we have used Equation (\ref{eq:dN1}).

Let $f(v)$ be the velocity distribution function, $N_Tf(v)\,dv=N_T g(\xi) \, d\xi$, where the left side is the number of clumps with velocities between $v$ and $v+dv$. Then,

\begin{equation}
    f(v)=g(\xi)\left| \frac{d\xi}{dv}\right|.
\end{equation}

Clearly, 
\begin{equation}
    \int_{v_{min}}^{v_{max}}f(v)\,dv=1
    \label{eq:normalize}
\end{equation}

From Equation~(\ref{eq:vm2}) we obtain 

\begin{equation}
    \left| \frac{dv}{d\xi}\right|=\frac{2\xi \xi_0 v_0}{(\xi^2+\xi_0^2)^{1/2}}=
    \frac{4v_0^2 \xi_0^2}{\xi v^3}
\end{equation}

to finally express $f(v)$ as 

\begin{equation}
    f(v)=\frac{4(3+\alpha)v_0^2 \xi_0^2}{\xi_c^{3+\alpha} v^3}I_\alpha
    \label{eq:distrfunct}
\end{equation}

where we have used Equation (\ref{eq:normalize}).

According to Equation (\ref{eq:vm2}) the distribution function for the velocity has a lower limit,

\begin{equation}
    v_{min}=\frac{2v_0}{(1+\xi^{'2}_c)^{1/2}}
    \label{eq:vmin}
\end{equation}

and an upper limit

\begin{equation}
    v_{max}=2v_0.
    \label{eq:vmax}
\end{equation}

Additionally, the velocity distribution given by Equation (\ref{eq:distrfunct}) has a maximum and/or a minimum given by the condition,

\begin{equation}
{\left[ \frac{dI_\alpha}{dv} \right]_{v_*} }=\frac{3I_a(v_*)}{v_*}
\end{equation}

for any of maximum or minimum at $v_{*}$.

We consider the following particular cases:

\begin{eqnarray}
    \alpha=0,  &  I=x_m \\
    \alpha=-1, \quad &  I={\rm{arcsinh}} (x_m/\xi) \\
    \alpha=-2, \quad  &  I=\frac{1}{\xi} {\rm arctan} (x_m/\xi) 
\end{eqnarray}

Then, a cluster with uniform density, i.e. $\alpha=0$, implies three different velocities that characterize the velocity distribution, $v_{min}$, $v_{max}$ and $v_{*,max}$ where

\begin{equation}
    v_{*,max}=\frac{4v_0}{\sqrt{3(1+\xi^{'2}_c)}}.
    \label{eq:vstar}
\end{equation}

So,

\begin{eqnarray}
    \frac{v_{*,max}}{v_{min}}=\frac{2}{\sqrt{3}}\approx 1.1547,\\
    f_*=\frac{9\sqrt{3}({1+\xi^{'2}_c})}{32\xi^{'3}_c v_0}
\end{eqnarray}
and
\begin{equation*}
    f(v_{max})=\frac{3}{2v_0\xi^{'2}_c}.
\end{equation*}
    
This velocity distribution is almost flattened in magnitude. Then, to explore the spatial distribution, numerical simulations are going to be performed and discussed in the next Section.


\section{N-Body simulations results}
\label{sec:simulations}

\subsection{Initial conditions}

In order to test the analytical analysis, we have performed several N-body
simulations of a particle of $M_*=10$ M$_\odot$ ``colliding" with a cluster of $N_T$ small mass particles to reproduce the velocity distributions derived in the last section. The free parameters of the simulations are $\alpha$, $v_0$ and the impact parameter of the massive particle respect to the cluster center {\it y*}.

We selected $\xi_c=0.67$ au as the cluster radius, which is a small region considering a proto-stellar envelope. However, a larger size would represent a weaker interaction where the internal forces should be considered. The number of total particles is $N_T=200$, that, approximately, have a typical separation of around $0.11$ au. The cluster is centered at the origin in the simulations and we consider a spatial distribution (the number of clumps per
unit volume) of the form,

\begin{equation}
    n(r) = \frac{(3+\alpha)N_T} {4 \pi \xi^{3+\alpha}_c} r^\alpha,
\end{equation}

where we analyze the cases $\alpha$= 0, for an homogeneous distribution, $\alpha=-1$ and $\alpha=-2$.

This cluster must be in dynamical equilibrium, then, assuming that every particle has a circular orbit around the distribution center of mass, the orbital velocity of a particle at a radius $r$ is, 

\begin{equation}
    v(r)=\sqrt{\frac{N_T M_i G}{\xi_c}} \left( \frac{r}{\xi_c} \right)^\frac{\alpha+2}{2},
\label{eq:orbit}
\end{equation}

where $M_i$ is their individual mass and $G$ is the gravitational constant. The analytical model considers the cluster particles at rest, so we choose $M_i=10^{-10}$M$_\odot$ to have a very small orbital velocity $\sim 10^{-2}$ km s$^{-1}$ that, effectively, allows us to use zero velocity for every cluster particle. Also, the force between two cluster particles is negligible. We use a random number generator that chooses a number $\eta$ uniformly distributed in the interval [0,1]. The value is
related to the radial distance as
\begin{equation}
    r=\xi_c \eta^{1/(3+\alpha)},
\end{equation}
from which we can sample $\xi$ as a function of the random number $\eta$ (for more detail see \citealt{RGETAL07}). We assigned random 
directions to the position vector of each particle.

We also included a single massive particle with a mass of 10~M$_\odot$, moving towards the particle distributions with $v_0=$100, 200, 300 and 400 km s$^{-1}$. At $t=0$, the massive particle starts moving from the Cartesian point (-10 au, {\it y*},0), in a direction parallel to the {\it x-axis} towards the clump distribution. Table \ref{tab:models} shows the initial velocity $v_0$ and the initial position {\it y*} over the {\it y-axis} of the massive particle and the exponent $\alpha$ of the distribution of particles at rest for the simulations presented in this paper.

In order to obtain a statistically significant result to compare with the velocity distribution, we have developed sets of 10 random distributions in each of our models.  An stability of our numerical solver is presented in the appendix.

\begin{table}[]
    \caption{Numerical simulation models of a particle with 10 M$_\odot$ moving at $v_0$ towards a cluster of number distribution with a parameter $\alpha$ from the point~(-10 au, {\it y*}, 0)}
    \centering 
    \begin{tabular}{|c| c| c| c|}
    \hline
        Models & $v_0$ &  $y_*$ & $\alpha$ \\
          & [km~s$^{-1}$] & & \\
        \hline
                \hline
        v200R0 &  200     &   0. &  0 \\
        v200R1 &  200     &   0.&  -1 \\
        v200R2 &  200      &   0.&  -2\\        
        v100R1 &  100      &   0.&  -1\\
        v300R1 &  300      &   0.&  -1\\    
        v400R1 &  400     &   0. &  -1 \\
\hline
        v200R1s05 &  200   & $\frac{1}{2}\xi_c$&  -1 \\
        v200R1s1 &  200  & $\xi_c$ &  -1 \\
                v200R0m &  200     &   0. &  0 \\
                        \hline
    \end{tabular}
   
    \label{tab:models}
\end{table}

\subsection{Comparison with analytical results}

\begin{figure}[ht]
\includegraphics[width=0.97\columnwidth]{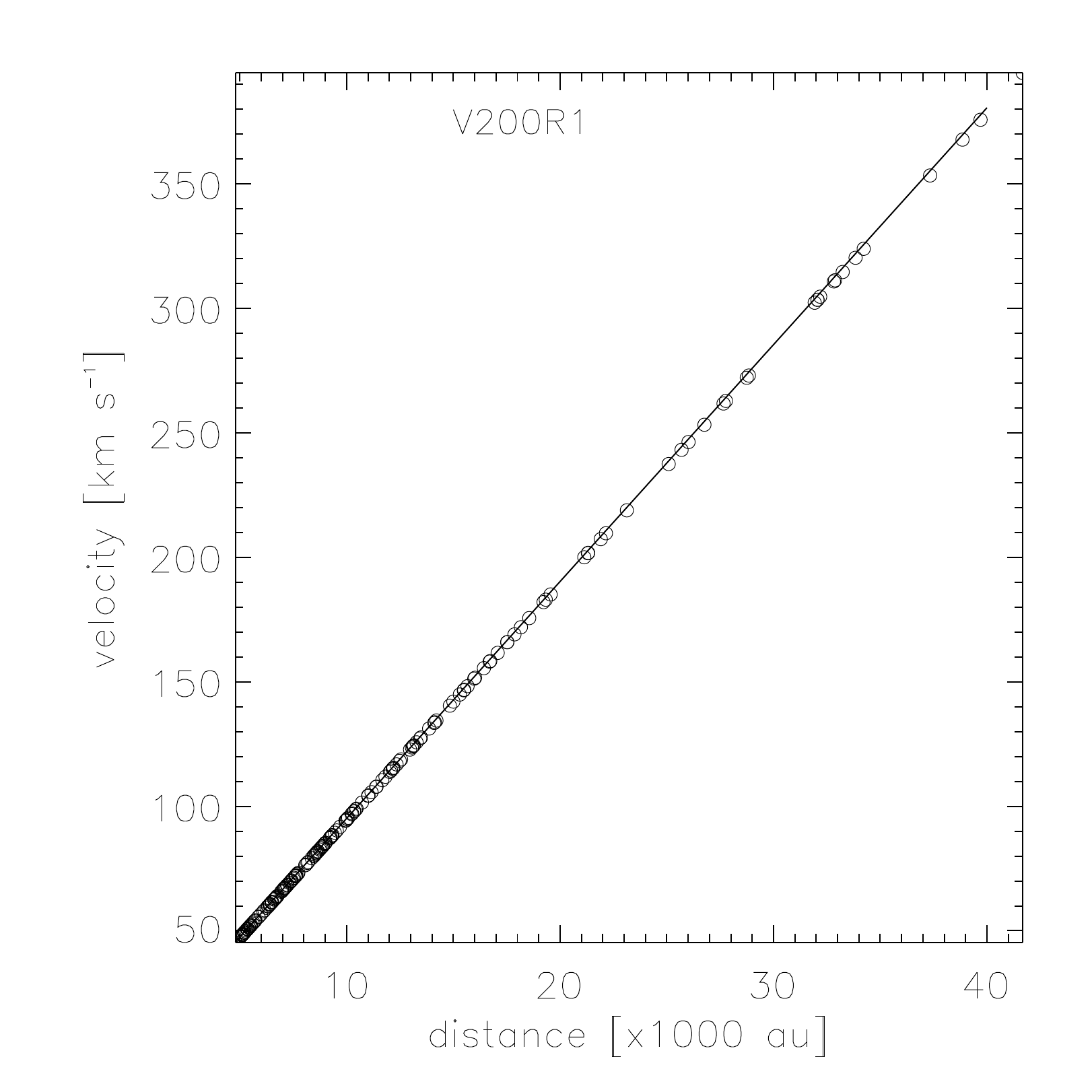}
\caption{The graphic represents the velocity $v_\infty$ (which is constant after the interaction) of each particle in the v200R1 model (open circles) related to the distance from the ejection point at evolutionary time of 500 yr. The solid line is a model of an homologous expansion ({\it Hubble law}) at time = 500 yr, and maximum distance of $4\times 10^4$ au.}
\label{fig:hubblelaw}
\end{figure}

 First of all, the consideration of negligible mass particles allowed us to ignore the potential energy of the cluster since the kinetic energy of the massive particle is greater by several orders of magnitude and. In this approximation the massive particle is not affected by the interaction. For the small mass particles, there is a very fast interaction that ejects them in almost all directions, Fig.~(\ref{fig:hubblelaw}) shows the velocity $v_\infty$, constant after the interaction, related to the distance, in au, at evolutionary time of 500~yr for the model v200R1 (open circles) and the homologous expansion model (solid line) with final distance of $4\times 10^4$ au at 500~yr. The average difference between the dynamical age of each point in the simulation and the model is $\sim 0.65$ yr, corresponding to the time when the low mass particles get to $v_\infty$.  As we can see, the numerical model follows a  {\it Hubble type law}, however, the small difference is due to: a) the time it takes for the massive particle to get the center of the particles cluster and get out of there, $\sim 0.25$ yr, considering $v_0=200$ km/s  and b) the time for the dynamic interaction between the low mass particles with the massive particle to get their terminal velocities ($\sim 0.4$ yr).

\begin{figure*}
\includegraphics[width=1.97\columnwidth]{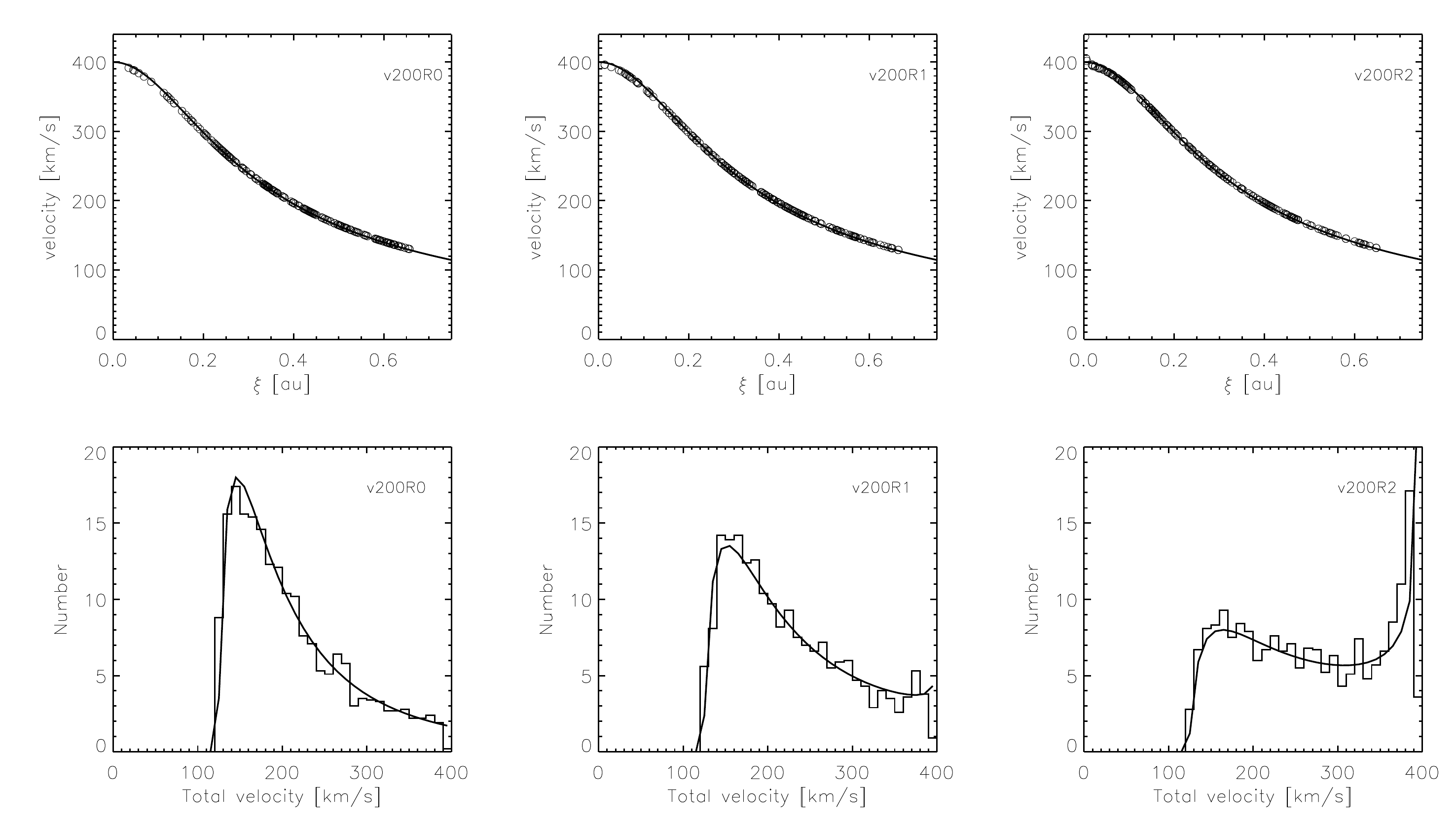}
\caption{Analytical and numerical results for models v200R0, v200R1 and v200R2, left, center and right columns, respectively. The velocity versus the initial impact parameters are plotted in the upper panels using open circles for the particles of the numerical simulation and a solid line for the analytical results, and the final velocity distributions are plotted in the bottom panels, using histograms for the numerical model results. The analytical solution (Sec.~\ref{sec:model}) is represented as a superimposed curve.}
\label{fig:v200rsdv10}
\end{figure*}

Fig.~(\ref{fig:v200rsdv10}) shows 
two rows, the upper panels show the velocity as a function of the impact parameter $\xi$, the solid line is the solution presented in the Equation~(\ref{eq:vm2}) and the open circles are each of the low massive particles from our numerical models. The lower panels, in the same figure, show the histograms obtained by integrating the distribution function (Equation~(\ref{eq:distrfunct})) for bins with a width $\Delta v=10$~km~s$^{-1}$,  in solid line, and the numerical solution is showed in bins. The models v200R0, v200R1 and v200R2, corresponding a models with a initial velocity of the massive particle of 200~km~s$^{-1}$ and particle distribution with $\alpha=0,-1$ and $-2$, left, center and right column, respectively, are presented in the columns of this figure. This figure (Fig.~(\ref{fig:v200rsdv10})) shows a very well agreement between the numerical and analytical results, the minimum velocity is given by Equation (\ref{eq:vmin}). For the models with $v_0=$200~km~s$^{-1}$ we use $\xi_c=3\xi_0$ (three times the gravitational radius), therefore $v_{min}=126.4$~km~s$^{-1}$, showed in this 3 models (v200R0, v200R1 and v200R2) independently of $\alpha$. The maximum velocity is given by equation (\ref{eq:vmax}), and in this case is $400$~km~s$^{-1}$ for particles with $\xi=0$, and the velocity of our numerical models tends to this value and it is more evident in the model with $\alpha=-2$, where the particles are more concentrate at
the center of the cluster, in this case, the velocity distribution is vertically asymptotic in $v_{max}$. In the case of constant density, $\alpha=0$, $v_{*,max}=$145.47~km~s$^{-1}$ and for this $\alpha$ the distribution has not a minimum (local or global). For the cases of $\alpha=-1$ there are no analytical { expressions} for the position of the maximum or minimum in the velocity distribution but this can be obtained semi-analytically, for the case of $\alpha=-1$, $v_{*,max}$=151.2~km~s$^{-1}$,  and $v_{*,min}$= 375.1~km~s$^{-1}$, and  for $\alpha=-2$, $v_{*,max}$=163.5~km~s$^{-1}$,  and $v_{*,min}$= 308~km~s$^{-1}$. These analytical values are in good agreement with the numerical ones showed in Fig.~(\ref{fig:v200rsdv10}).

\begin{figure*}
\includegraphics[width=1.97\columnwidth]{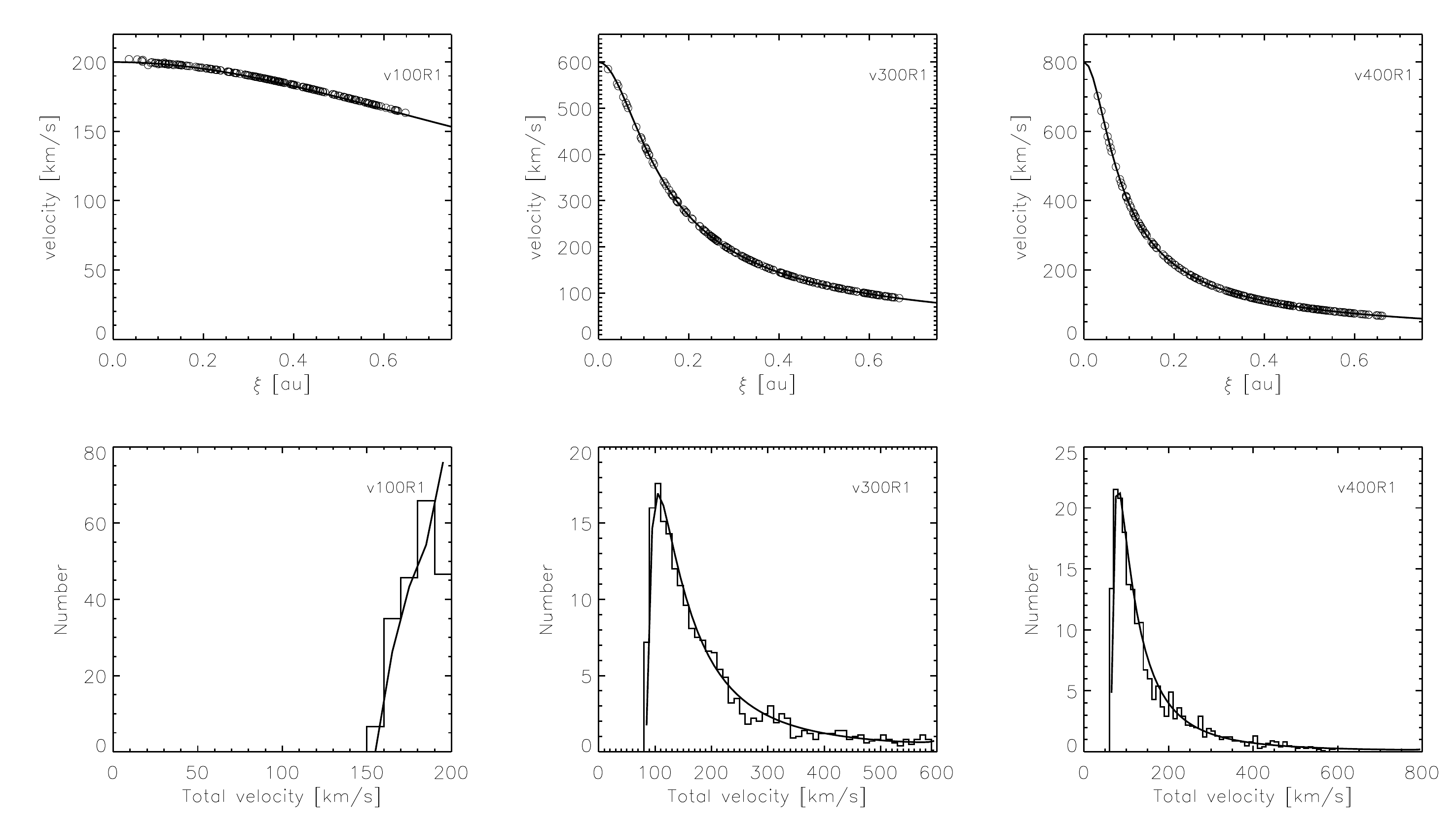}
\caption{Analytical and numerical results for models V100R1, V300R1 and V400R1, left, center and right columns, respectively. The velocity versus the initial impact parameters are plotted in the upper panels using a open circles for the particles of the numerical simulation and a solid line for the analytical results, and the final velocity distributions are plotted in the bottom panels, using histograms for the numerical model results. The analytical solution (Sec.~\ref{sec:model}) is represented as a superimposed curve.}
\label{fig:vsr1dv10}
\end{figure*}

Moreover, we ran models with different velocities, holding the $M_*$ and the $\alpha=-1$ fixed, therefore the gravitational radius is different in each of our models, $\xi_0=0.85$, $0.22$, $0.095$ and $0.053$ au, for v100R1, 
 v200R1, v300R1 and v400R1, respectively. These models, with the exception of the v200R1 are showing in Fig.~(\ref{fig:vsr1dv10}). The description of the rows (and the plots in them) is the same as Fig. (\ref{fig:v200rsdv10}) except that the column, in this figure, corresponds to the models V100R1, V300R1 and V400R1, left, center and the right column, respectively. v200R1 is in the second { column} of Fig.~(\ref{fig:v200rsdv10}). As well as the previous figure, the analytical solution is in accordance with the numerical solution. For all these models we ran the numerical simulations using $\xi_c$=0.66 au, which is 3 times the gravitational radius of models with $v_0$=200~km~s$^{-1}$. As one expects, the particles of the model V100R1 are distributed in a cluster with a radius smaller than its gravitational one. Therefore the velocity distribution have values around the maximum speed, in this model $v_{max}=$200~km~s$^{-1}$. In our models with higher initial velocities, which means lower gravitational radius, the cluster is distributed in a larger volume, because we have fixed the radius of the cluster, and the initial impact parameters of the particles are larger, so that the velocity distribution has a similar form but distributed in a larger range of velocities (corresponding at maximum velocity on each model).

\subsection{Non-symmetrical collision}\label{sec:cross}
\begin{figure}
\includegraphics[trim=35 4 10 15,clip,width=\columnwidth]{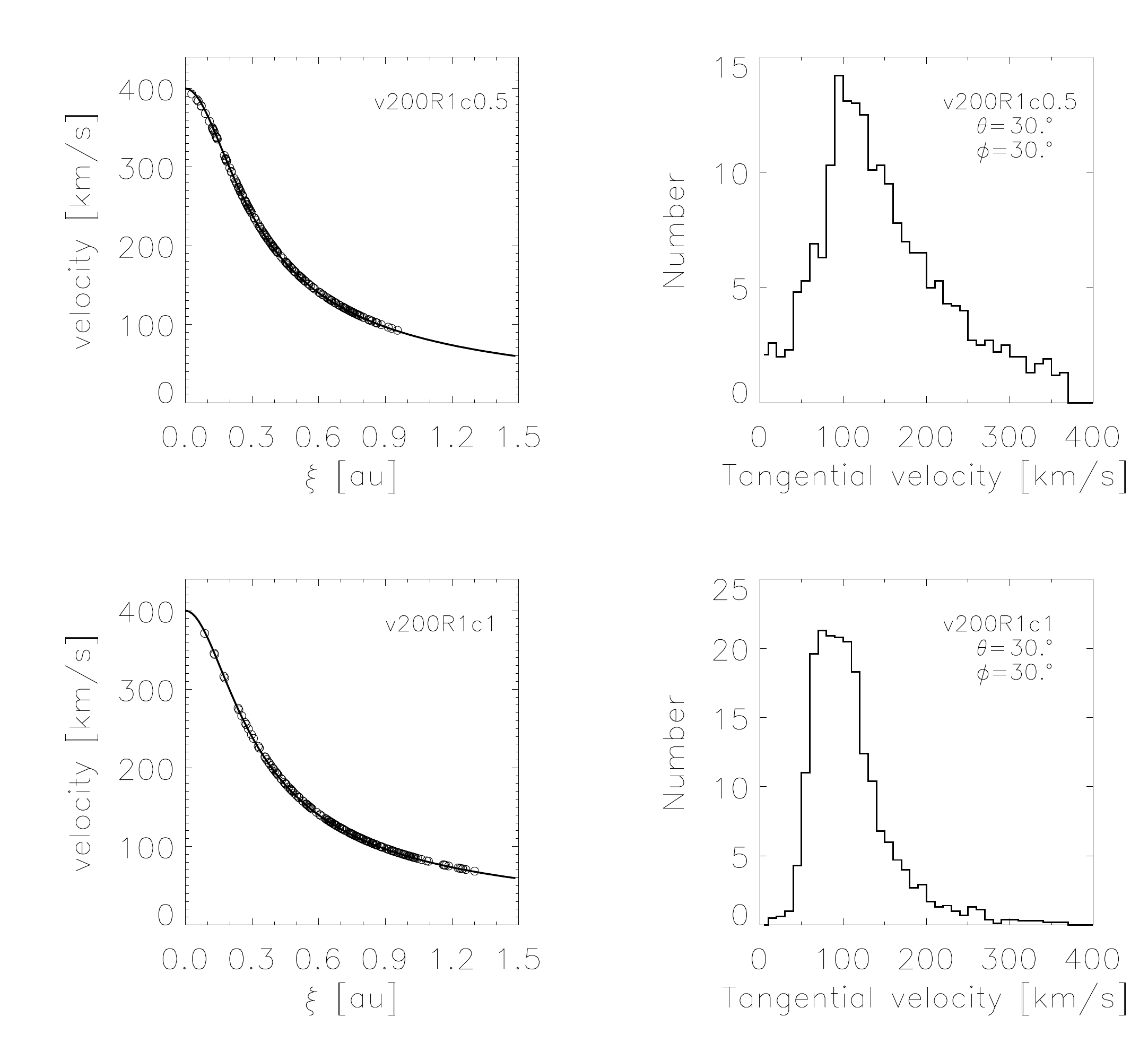}
\caption{Velocities distribution of the model v200R1s05 and v200R1s1, upper and bottom panels respectively. The velocity as a function of initial impact parameter (left column) and the final velocity distribution (right panels) with description same as Fig. (\ref{fig:v200_dv10}.)}
\label{fig:v200r1dv10xis}
\end{figure}

It is also important to analyze the distribution of the tangential velocities when the massive particle does not pass
through the center of the particles distribution. In the models v200R1s05 and v200R1s1, the massive particle goes through the cluster distribution  at half the radius of the cluster on the $y$-axis and on the edge of the cluster, on the $y$-axis as well.  Fig. 
(\ref{fig:v200r1dv10xis}) shows the model v200R1s05 in the upper row panels, and v200R1s1 in lower row panels, where the right column shows the velocity as a function of the impact parameter and the left column shows the histogram of the tangential velocities. As one can see the velocities of the particles of the model v200R1s1 are minors, because the impact parameter of the particles are larger than the v200R1s05's particles, or for the case where the massive particle crosses the center of the distribution where the particle density is greater, because it increases towards the center of the cluster. The velocity in the plane of the sky the plots have maximum in about 100~km~s$^{-1}$ and 90~km~s$^{-1}$ for v200R1s05 and   v200R1s1, respectively (see Table \ref{tab:models}). 

\section{The case of Orion BN/KL}
\label{sec:Orion}

Several examples of explosions generated by dynamical interaction are presented in the astrophysics literature, i.e. globular clusters disintegration, destruction of planetary systems in dynamical encounters (\citet{SPETAL09}) or in a star formation explosive region such as Orion BN/KL. In Orion BN/KL there are around 200 clumps moving into the interstellar medium with actual velocities between  100 and 300~km~s$^{-1}$. They are moving away from a common origin with a set of protostars that apparently interacted in the past, following a homologous expansion law, better known as {\it Hubble Law}. RO19~(a and b) calculated the initial velocity distribution of the clumps, considering their deceleration due to its interaction with the interstellar medium. In Fig. 7 from RO19b, they showed an initial velocity distribution of the clumps in Orion BN/KL with two maximum values, in $\sim$200 and 400~km~s$^{-1}$, global and local maximum respectively.  In this section, we are using our dynamical interaction model to propose an ejection mechanism to account for the explosive properties of the Orion BN/KL outflow.  In this section, we are going to analyze the importance of the projection angles and the mass of the clumps in the final velocity distributions and also, we are going to compare it with the observational data.

\subsection{Projection angles} \label{sec:proj}
Using the results of the model v200R1 we are obtaining the projection, on the plane of the sky, of the position and velocity for each particle in our models.  To achieve this, we  must take into account the rotation projections on the plane of the sky, to get the projected positions and velocities for each one of the particles.  We use the rotation matrices $R_x$, $R_z$, which are rotations with an angle $\theta$ and $\phi$ in the $x$ and $z$ axes.

Using the projected velocity, one can calculated the tangential velocities (the velocities on the plane of the sky),
\begin{equation}
    v_t=\sqrt{v^2_{tot}-v^2_{r}},
        \label{eq:vtang}
\end{equation}
with, $v_{tot}=\sqrt{v^2_x+v^2_y+v^2_z}$ and the radial velocities of each of the particle is given by the projected $z$-velocity,
\begin{equation}
    v_r=v'_{z}.
    \label{eq:vradial}
\end{equation}

For this analysis, we present the result using the rotation angles $\theta=30^{\circ}$ and $\phi=30 ^{\circ}$. These angles are selected in order to obtain a wide tangential velocity distribution with a maximum value in around 200~km~s$^{-1}$, like the case of Fig. (7) in RO19b. It is important to note that it is not the aim of this paper to explore in detail the combination of angles and/or the precise initial clump distribution that reproduce the initial velocities proposed for this object. In that case, one must consider other effects, such as the dynamics of gas, but we are interested in proposing this kind of explosion as a possible mechanism for the formation of this type of object. 

\begin{figure}
\includegraphics[trim=40 4 10 15,clip,width=\columnwidth]{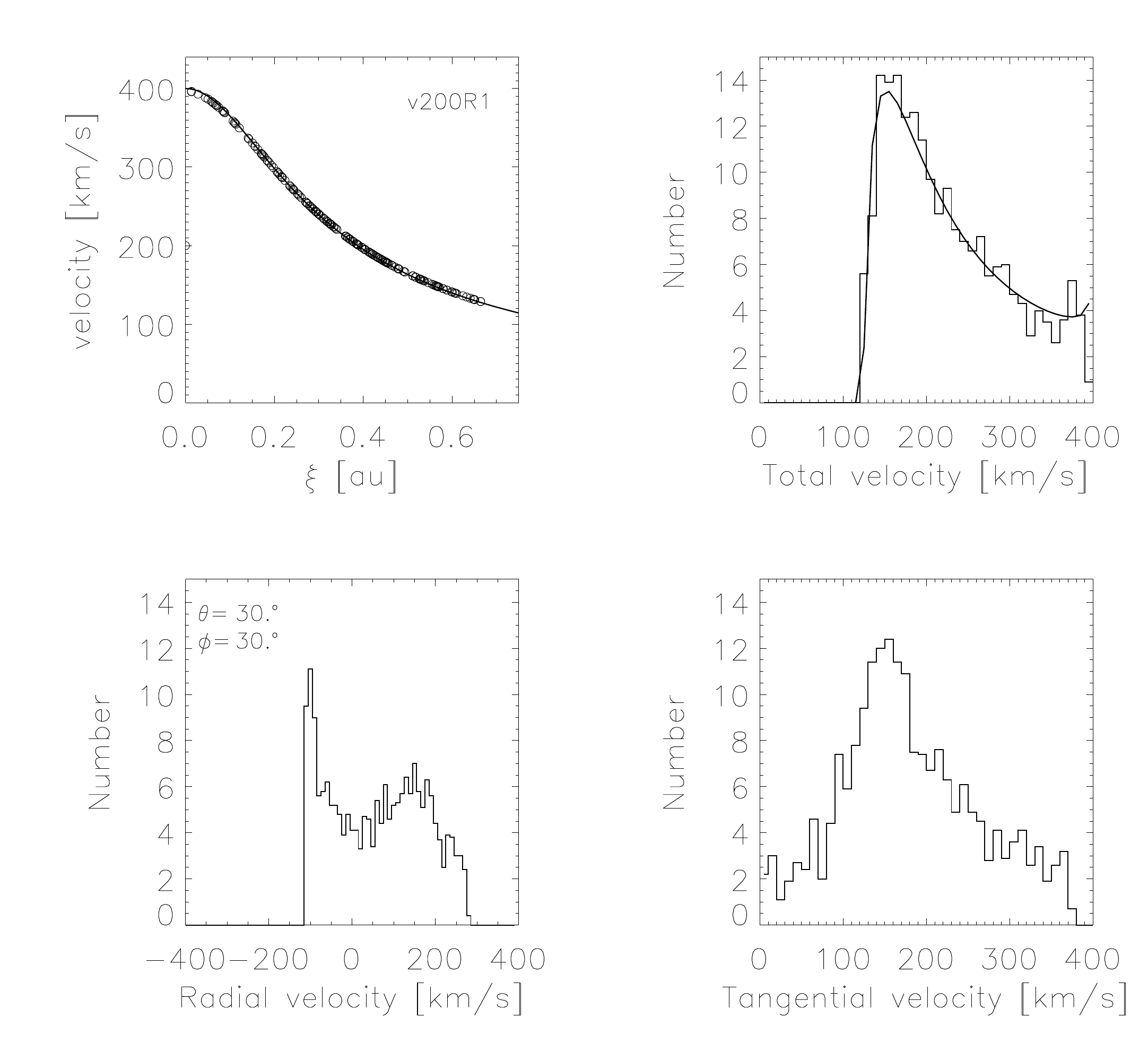}
\caption{Velocities distribution of the model v200R1. The upper panels are the velocity as a function of initial impact parameter and the final velocity distribution with description same as Fig. (\ref{fig:v200_dv10}). The bottom panels are the radial velocity (bottom left panel) using $\theta=30^{\circ}$ and $\phi=30^{\circ}$ and the velocity in the plane of the sky (left bottom panel).} 
\label{fig:v200_dv10}
\end{figure}

Fig. (\ref{fig:v200_dv10}) shows the velocity as a function of distance (upper left panel), the total velocity, radial velocity, and tangential velocity distributions (upper right, lower left and right panels, respectively). 

  Using  the Equation (\ref{eq:vradial}), we have  calculated the radial velocity distribution in a range of the velocities between about -110 and 270~km~s$^{-1}$, with maximum values in about -100  and 150~km~s$^{-1}$. Using these radial velocities in Equation (\ref{eq:vtang}), we obtained the velocity on the sky plane for each of the particle in the set of numerical simulation called v200R1. The tangential velocity distribution of this model has a maximum value in about 150~km~s$^{-1}$ and the shape of this tangential velocity distribution is similar to the initial velocity distribution presented in Fig. (7) of RO19b, even when the distribution is spread over a large velocity range. 

\subsection{The massive clumps}

 However, the mass of clumps in the region of Orion BN/KL, can be estimated by using the total mass of the moving gas in the region and, for simplicity, dividing it by the total number of current observed clumps. Additionally, RO19b predicted the initial mass of the clumps (see fig. 6 of that paper). The initial mass of each individual clump is around 10$^{-2}$~M$\odot$, and the total mass of these particles is comparable with the mass of the more massive particle (the star mass, i.e. 10 M$_\odot$). The effects of the interaction between the low-mass particles, and their contribution in the global motion of this event are not considered in the first models presented in this work. Nevertheless, the high velocity of the massive particle, and its momentum, play a more important role than the mass of the particles. In order to prove it, we ran a final model, v200R0m, where each of the low-mass particles have a 0.01~M$_\odot$ mass. We have assigned a random direction for the orbital velocity, assuming circular motion according to Equation (\ref{eq:orbit}) and they are in quasi-equilibrium with each other, while the massive particle collides this cluster and interacts with them. We have calculated the random positions and  orbit directions of each of them following this procedure:
\begin{enumerate}
    \item for each particle, assign a random radius using,
    \begin{equation}
        r=\xi_c \eta^{1/(\alpha+3)}
    \end{equation}
    \noindent where, $\eta$ is uniform random number between 0~$\to$~1.
    \item we can calculate the cartesian coordinates, $x$, $y$ and $z$, using
    \begin{equation}
         x=r[\sin(\theta)\cos(\phi)],     \notag
     \end{equation} 
     \begin{equation}
         y=r[\sin(\theta)\sin(\phi)], \notag
      \end{equation}
    \begin{equation}        
         z=r[\cos(\theta)]
    \end{equation}
   \noindent where, $\theta=\arccos(2\eta_t-1)$ and $\phi=2\pi\eta_p$, and $\eta_t$ and $\eta_p$ are the uniform random numbers between 0 $\to$1.
    \item we assigned an orbital velocity, $v_r$, using Equation (\ref{eq:orbit}),
    \item we calculated $v_x$, $v_y$ and $v_z$ using,
        \begin{equation}
        v_x=  v_r[\cos(\phi)\cos(\theta)\sin(\chi)+\sin(\phi)\cos(\chi)]\notag
        \end{equation}
        \begin{equation}
        v_y= v_r[\cos(\theta)\sin(\phi)\sin(\chi)-\cos(\phi)\cos(\chi)] \notag
                \end{equation}
                \begin{equation}
        v_z=-v_r\sin(\theta)\sin(\chi)
    \end{equation}
\noindent where, $\chi= 2\pi \eta_x$ and $\eta_x$ is a uniform random number between 0 $\to$ 1.

\end{enumerate}

\begin{figure}
\centering
\includegraphics[trim=40 4 10 15,clip,width=0.8\columnwidth]{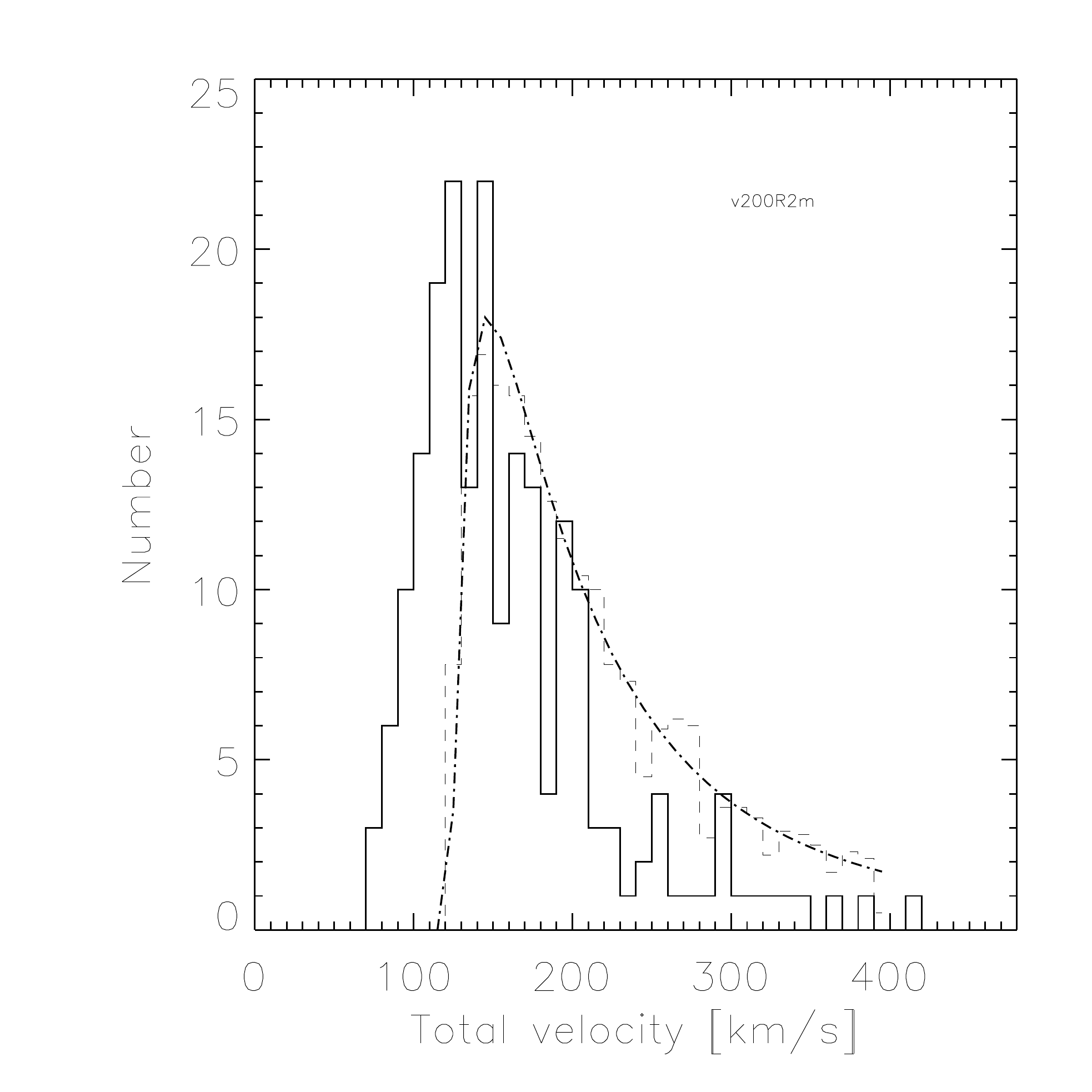}
\caption{Velocities distribution of the model v200R0m. The solid and dashed lines are the results of the model v200R0m and v200R0 and the dash-dotted line is the analytical result obtained in the Sec.~\ref{sec:model}.} 
\label{fig:v200m}
\end{figure}

 Similar {\bf to} previous models, we have run a set of 10 simulations using different random  distributions. In this model we used a uniform distribution and the massive particle has an initial velocity of 200 km/s. Fig.~\ref{fig:v200m} shows the histogram of the total velocity for the model v200R0m in solid line, the model v200R0 (with clumps of negligible mass) in dashed line, and the dash-dotted line is the analytical solution (Section~\ref{sec:model}). The shape of these histograms is similar, but the model v200R0m present the maximum of the distribution in a small velocity, about 130 km/s, 20 km/s lower than the model v200R0. The minimum velocity of any particle is about 80 km/s, it is also 30 km/s lower than the model v200R0, and also the particle with higher velocity is larger than in the model where the low-mass of the particles are negligible. The maximum velocity obtained in our numerical simulation is very similar at the maximum velocity predicted by RO19b (see fig. 7 in that paper), but the shape of the velocity distribution does not fully agree with the predicted in RO19b. However, as shown in the v200R1c0.5 and v200R1c1 models (Fig.~(\ref{fig:v200r1dv10xis})), the projected velocity distribution is also a function of the position through which the massive particle cross the initial particle distribution (\S~\ref{sec:cross}) and the projection angle  which the observed velocities are calculated (\S~\ref{sec:proj}). However, a study of these parameters, for this particular object, is outside the scope of this paper and will be addressed in subsequent works. 

\begin{figure}
    \centering\includegraphics[width=7.3cm]{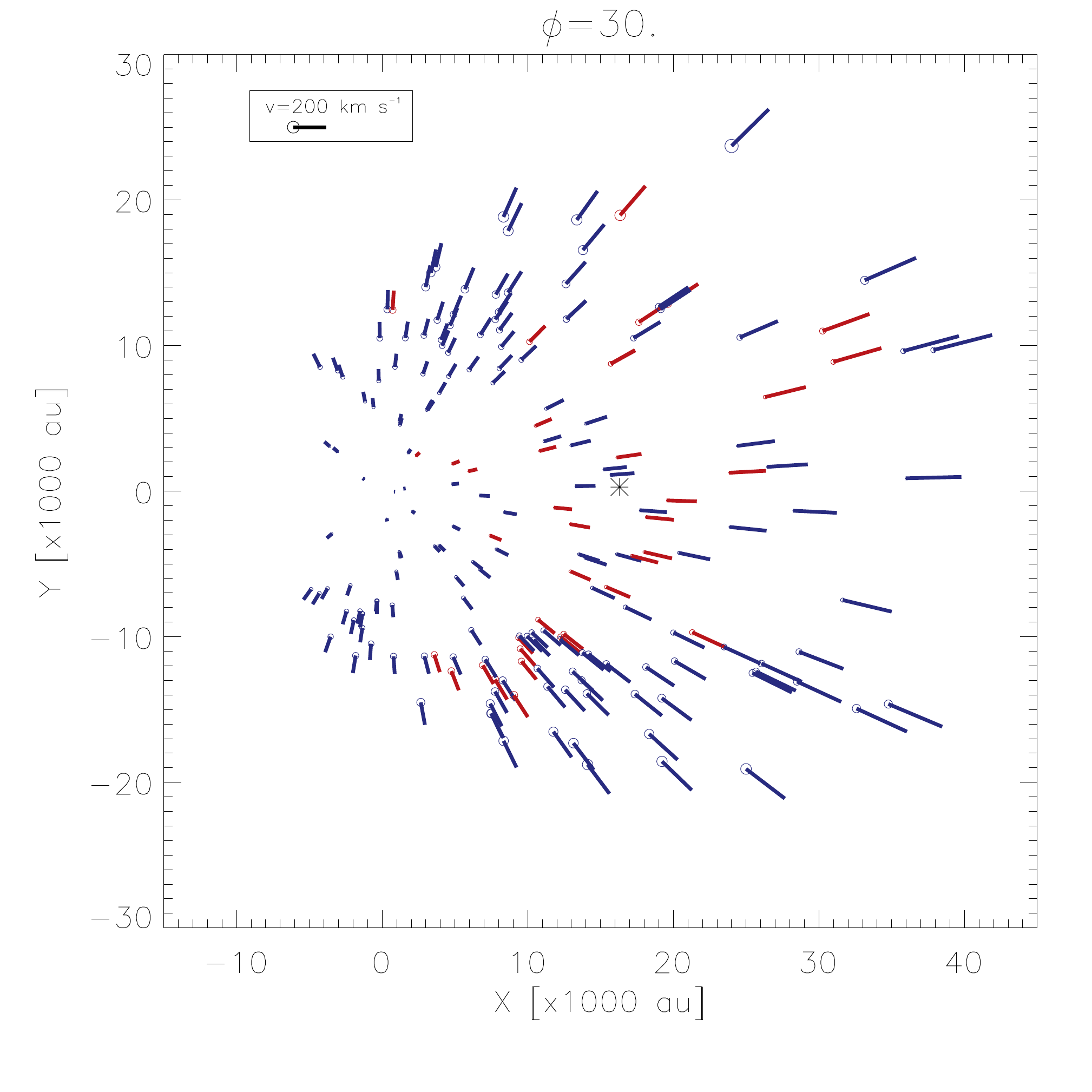}
%
    \centering\includegraphics[width=7.3cm]{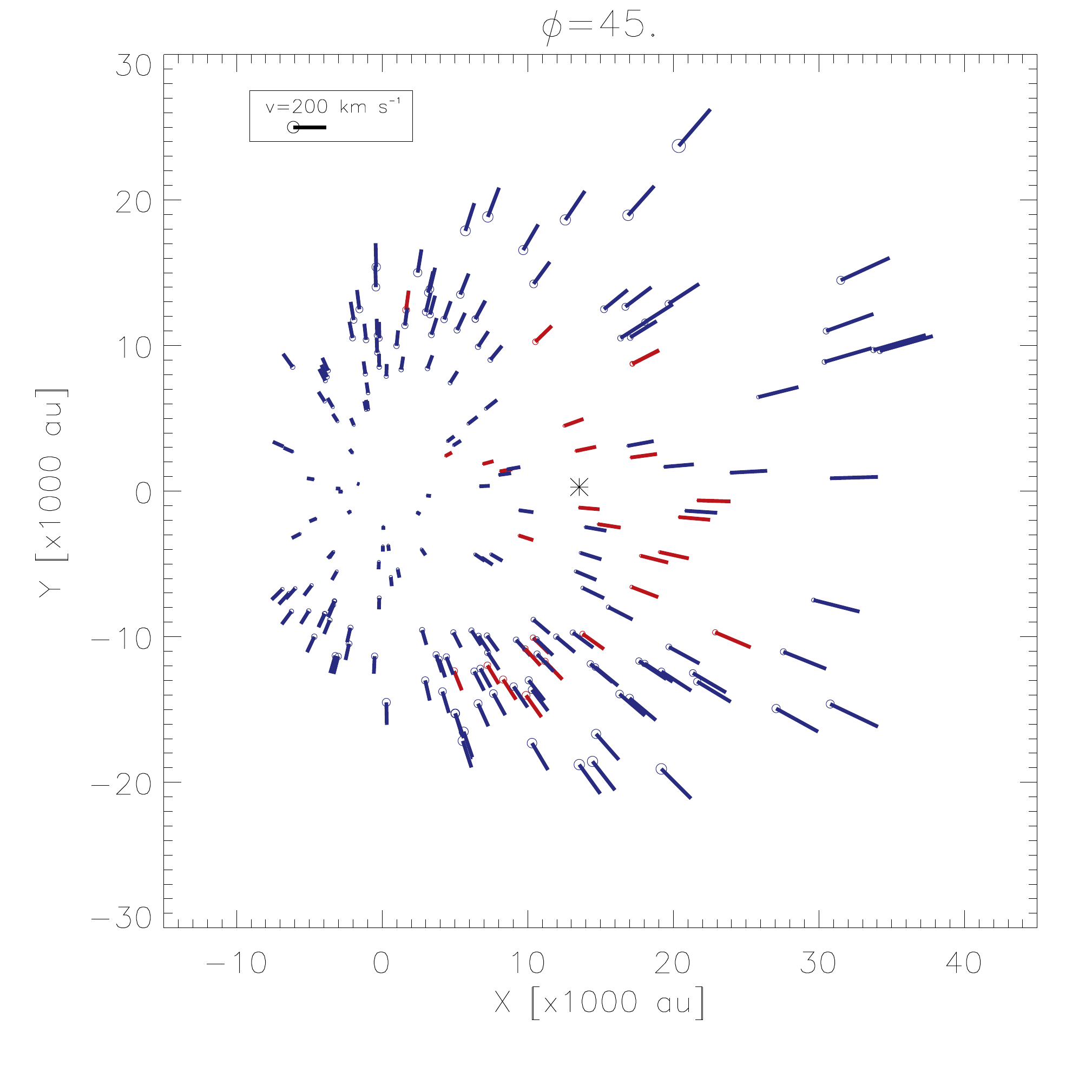}
    \centering\includegraphics[width=7.3cm]{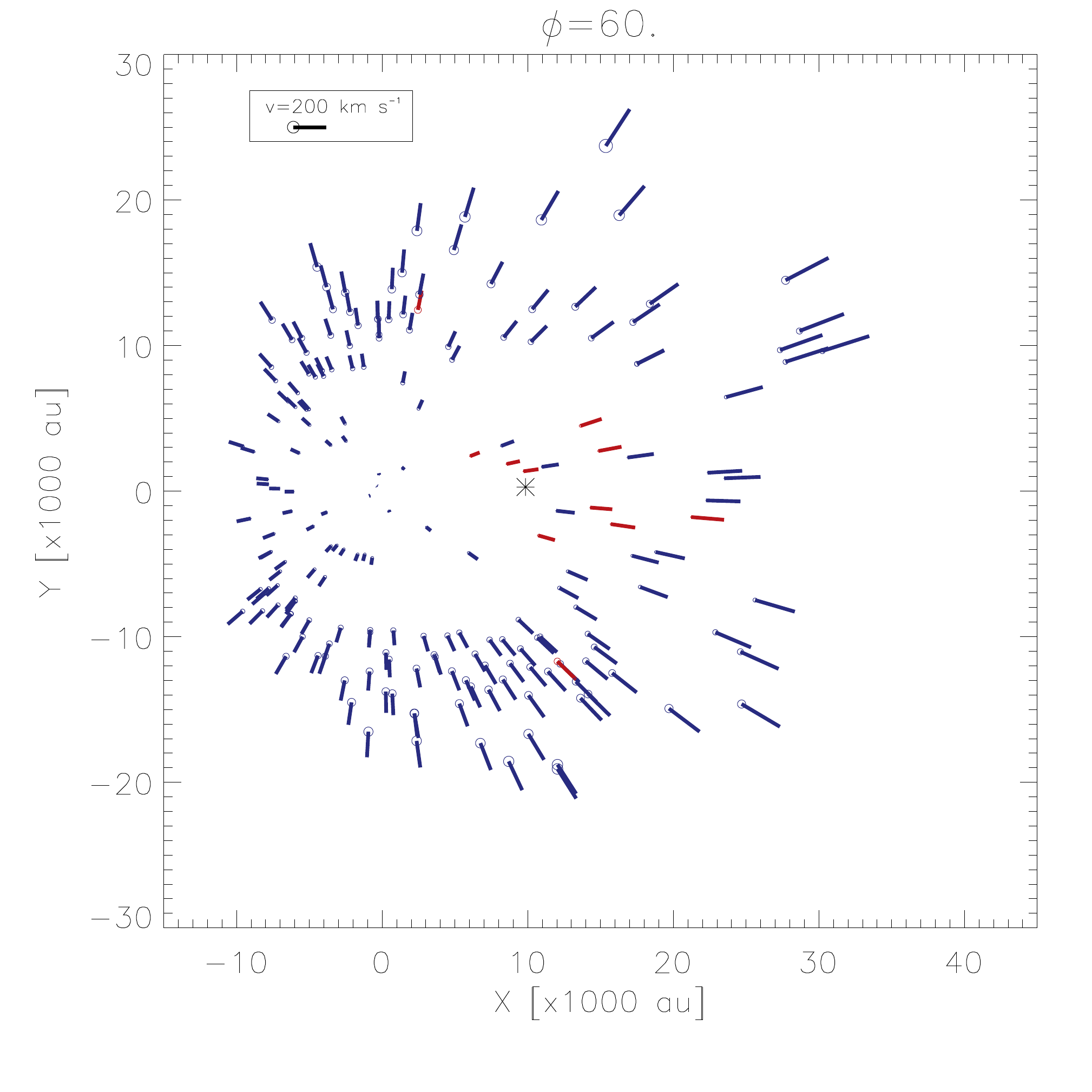}
    \caption{Position and velocities for each of the low massive particles for one the numerical model of the v200R1 model. The upper, middle and bottom panel are the projected position of the each particles for $\phi=30$, $45$ and $60^\circ$. The blue and red arrow are shown the radial velocity direction, moving or moving away from us, respectively. The size of the arrows are the tangential velocity.}
 \label{fig:posandvel}
\end{figure}

Finally, Fig. (\ref{fig:posandvel}) shows the position and velocities of a single simulation of the model  v200R0m , with projection angles of $\theta=0$ and $\phi=30, 45$ and $60$, for upper, middle and bottom panel respectively. In this figure we plotted the tangential velocity of all the particles using red or blue arrows for positive and negative radial velocities respectively. We also present, in the upper panel, the position of each particle when they  crossed the y-axis (where they were blown away by dynamic interaction, according to Equation (\ref{eq:r-axis})). As one can see the particles are ejected from a very small volume (about 0.3 au) that is insignificant with the size of the event after 500 years  (about $4\times 10^4$ au) and is in accordance with the observations that suggest a single ejection point at least with the current resolution. Thus, these types of explosions seem to be ejected in a singular place in space, as well as the Orion BN/KL event. 

 However, the number of particles approaching or moving away from us is dependent of the projected angle, but the explosion is not isotropic, at all, in the x-axis which is the axis of movement of the massive particle.  But the morphology of the explosion is similar to the observed in Orion BN/KL. It is important to note that the massive particle in the Orion BN/KL explosion should be, at least, a runaway massive star with a velocity of 150~km~s$^{-1}$, but the dynamical interaction of the massive star with the clumps with a total mass comparable to the star mass could decrease substantially the velocity of the star, at the end of the interaction. However, it is not the goal of this work to study the dynamic interaction of low-mass particles.

\section{Conclusion}
\label{sec:conclusions}

We presented analytical and numerical solution of the dispersion of the particles because of the dynamical interaction with a single massive particle. We have considered that the particles are seating into a cluster, and they  have a negligible mass with respect to the massive particle. The  dynamical interaction with the massive particle produces a quasi-isotropic ejection of the particles. 

Then, we carried out a set of numerical simulations of spherical distribution of mass-less particles (N-body simulations) for verifying our analytical solutions and obtaining an observational result and we have obtained a very good agreement between the numerical and analytical results.

The main conclusions are:
\begin{enumerate}

\item The gravitational ejection mechanism is able to accelerate small clumps or cloudlets (i.e. low mass gas fragments) to jet-like velocities and, therefore, it should be deeply explored in future work.

\item The terminal velocity $v_\infty$ of each particle is function of  its own impact parameter. 

\item The maximum terminal velocity is given by the limit when the particle has an impact parameter equal zero and it is two times the velocity of the massive particle ($v_{max}=2 v_0)$. The minimal terminal velocity,  is related with the cluster radius.

\item The ejection angle of each of the particles is linearly related with the terminal velocity, and 
therefore is related to the impact parameter. Compact distributions of particles that are dynamically disturbed by a massive particle, that passes through the center of the distribution, produce more collimated ejections than in the case of more scattered clusters.

\item The resulting dispersion has an explosive signature, such as: a) a small scale common origin, b) an isotropic distribution and c) velocities proportional to distance to that common origin.  This is the result of a short time interaction, which could be the mechanism that produced the explosive outflows. 

\item The distribution of the ejection velocities is a function of the exponent of the initial distribution of particles ($\alpha$), the gravitational and cluster radius, the massive particle velocity and the terminal velocity of each of the particles. The { minima and maxima}, local or global, in these distributions can be obtained analytically.

\item The off-center dynamical interaction produces a wider velocity distribution, and with smaller velocities.

\end{enumerate} 

 Then, a dynamical interaction between a massive object with a cluster of less massive particles is able to increase the individual energy of the cluster members producing an explosive event, instead of forming a new cluster with a massive particle or letting the massive particle to cross the cluster with a minor perturbation.

Finally,   we considered that the Orion BN/KL ejection was generated by a dynamical interaction. To demonstrate this, we have run a set of simulations where the true mass of gas in the fingers in Orion has been considered. Our models show that the interaction of a massive particle with a distribution of particles with the same mass as that observed in the Orion Fingers BN/KL produces an ejection of material in all directions and with a velocity distribution comparable to those observed in this region. Certainly a study of parameters, mass distribution of the clumps, impact parameters of the massive star, morphology of the cluster of clumps, etc., in addition to the projection angles which they are observed are parameters that must be explored in detail in a future job.

\begin{acknowledgments}
This project has received funding from the European Research Council (ERC) under the European Union's Horizon 2020 research and innovation programme, for the Project “The Dawn of Organic Chemistry” (DOC), grant agreement No 741002. We acknowledge support from PAPIIT-UNAM grants IN-109518 and IG-100218. The authors acknowledge Bertrand Lefloch for his useful comments to improve this work.  
\end{acknowledgments}



\appendix
\section{N-body's solver and stability}
\begin{figure*}[h]
    \centering\includegraphics[width=\columnwidth]{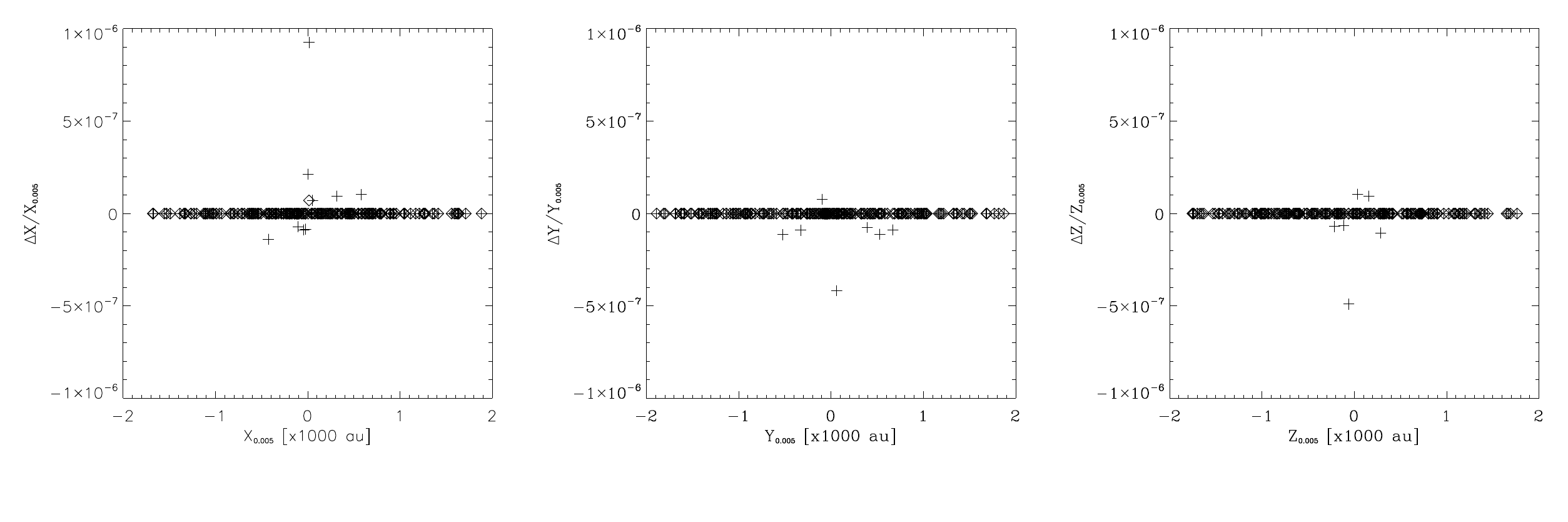}

    \caption{Relative differences between models with different $A$ constant, 0.5, 0.05, and 0.005, (see Equation (\ref{eq:dt})). We have calculated the difference of each of the particles with respect to the smaller timestep model (A=0.005). The left, center and right panels are the differences in X, Y and Z coordinates. The plus and diamond symbols are the model with A=0.5 an A=0.05, respectively.}
 \label{fig:restest}
\end{figure*}

{ The numerical method is a symmetrized leapfrog integrator with a variable timestep formalism, which is second order accurate and is able to preserve energy.} For the N-body solution, we {\bf have considered} $N$ particles with masses $m_i$, and position given by $x_i$, $y_i$ and $z_i$. The force between a pair of particles produce an acceleration, and the new position of each of the particles is strongly dependent of the time step $\Delta t$. A very large time step would solve incorrect trajectories and a small time step reproduces the real trajectory of each particle, dramatically increasing the computation time. In order to have an appropriate time step, we used a time step as, 
\begin{equation}
    \Delta t=A*\sqrt{\frac{R_{min}}{a_{max}}} 
    \label{eq:dt}
\end{equation}
where, $A$ is a constant between $0 \, \to \, 1$, R$_{min}$ is the mean distance between a pair of particles, and a$_{max}$, is the maximum acceleration of a single particle. In order to prove the solutions of our N-body solver, during the simulation time, we used a single particles distribution of the model v200R0, and we carry out the N-Body simulation, using different values of A, 0.005, 0.05, 0.5. Fig.~\ref{fig:restest} shows the relative position, $\Delta$X, $\Delta$Y and $\Delta$Z (for left, center and right, panels respectively), at evolutionary time =100 yr, with respect of the position of the model with smaller A value (A=0.005), where $\Delta$X=(X$_A$-X$_{0.005}$)/X$_{0.005}$, as for the others coordinates (Y and Z). The plus symbols are used for the model with A=0.5 and diamond symbols are used for the results of the model with A=0.05. The plots range (in the vertical axis) is -1$\times$10$^{-6}$ $\to$ 1$\times$10$^{-6}$, 
being, then, one thousandth percent the largest difference between the model A=0.5 and 0.005.  The small difference between models with different timestep added to the convergence with the theoretical results, guarantee that the code adequately solves the system of equations for the models used in this job.

\end{document}